# Explanation of the recent results on photoionization of endohedral atoms

**M. Ya. Amusia[1,2], L. V. Chernysheva[2] and E. G. Drukarev[3]**

[1]*The Racah Institute of Physics, the Hebrew University of Jerusalem, Jerusalem 91904, Israel*
[2]*A. F. Ioffe Physical-Technical Institute, St. Petersburg 194021, Russian Federation*
[3]*B. P. Konstantinov Petersburg Nuclear Physics Institute, Gatchina, St. Petersburg 188300, Russian Federation*

**Abstract:**

We suggest an explanation of the recently observed discrepancy between the experimental and theoretical results on ionization of atoms, encapsulated into the fullerenes by photons with the energies of about 80-190$eV$. On the ground of previous theoretical considerations we conclude that the photoionization of the caged atom without excitation of the fullerene shell has a low probability.

PACS numbers: 32.80Fb, 33.80Eh, 36.40.Cg

**1**. In recent publications on ionization of atoms caged inside the fullerene shell by photons, carrying the energies of about $\omega \leq 190 eV$ [1-3], the observed cross sections were compared with those, calculated for the isolated atoms [4-6]. The results of calculations strongly overestimated the experimental data. This is a direct manifestation of a very low probability of photoionization without an inelastic process in the fullerenes shell (FS) at relatively high photon energies, which was predicted in [7, 8]. As "inelastic" we name the processes, in which photoionization of the caged atom is accompanied by emission of an atom or an electron, excitation of collective states from FS, etc.

In the present Letter we compare the data obtained in [1-3] with our calculations. We demonstrate that in fact the inelastic processes become decisively important starting already from relatively low photoelectron energies, of about 80-100 eV.

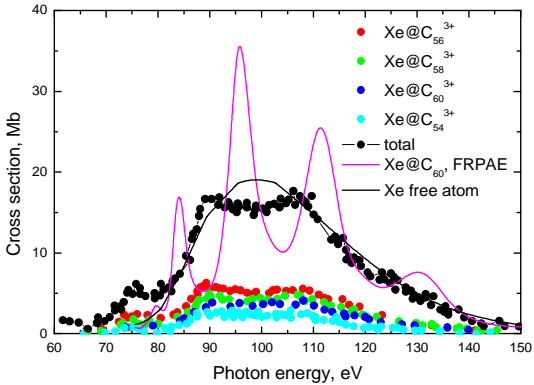

Fig.1. Total and partial photoionization cross-sections of $Xe@C_{60}^+$ compared to total cross-section for Xe. Experiment for $Xe@C_{60}^+$ – from [2], calculations – [4, 5]. The calculation cross-sections are multiplied by the reduction factor 0.62.

**2.** The results on photoionization of the 4$d$ subshell of xenon placed inside the fullerenes $C_{60}^+$ usually denoted as $Xe@C_{60}^+$ were reported initially in [1]. Much more detailed and accurate data on the same subject are presented in [2]. The paper [3] reports similar results for $Ce@C_{82}^+$. The experimental total photoionization cross-sections from [2] are presented in Fig.1 along with the results of calculations in frame of RPAE [4] multiplied by reduction factors, chosen to fit the sum rule[1]:

$$\frac{c}{2\pi^2}\int_I^\infty \sigma(\omega)d\omega = N \qquad (1)$$

Here $\sigma(\omega)$ is the total photoionization cross section, $I$ is the ionization potential, $N$ is the number of electrons in the ionized object. Expression (1) can be applied also to ionization of a single subshell. Using the experimental data on the total photoionization cross-section the authors of [1] found $N = 6.2 \pm 1.4$ for $4d$ - subshell in $Xe@C_{60}^+$. Note that for an isolated Xe atom $N_{4dXe} \cong 10$. So, the reduction factor is 0.62.

Multiplying the calculated results by the reduction factor 0.62 we find the experimental results for $Xe@C_{60}^+$ to become quite close to those of calculations for isolated Xe atom in the region $\omega \leq 150 eV$. However, we are confident and see some indications of this in Fig. 1 that even after this fitting the calculations carried out in various approaches ([4-6] and references therein) still overestimate experimental data at the upper end of the interval, i.e. at $140 eV \leq \omega \leq 150 eV$.

From our point of view the most essential result of [1] is the measurement of photoionization cross section of the caged atom in $Xe@C_{60}^+$ that is accompanied by emission of carbon atoms. It is seen from Fig. 1 that the cross-sections of ionization to considered final states $Xe@C_{56}^{3+}$, $Xe@C_{58}^{3+}$, $Xe@C_{60}^{3+}$, and $Xe@C_{54}^{3+}$ all are important. We see also that the channel $Xe@C_{60}^{3+}$ contributes only $1.71 \pm 0.38$ to the left hand side of Eq. (1). This means that the integrated probability $P = \int_I^{\omega_{max}} P(\omega)d\omega$, where $\omega_{max} \simeq 150 eV$, of the "elastic" photoionization of $Xe@C_{60}^{3+}$ (i.e. that, in which the fullerenes shell remain in the ground state) is $P \leq 20\%$. Note that processes with emission of several electrons were not taken into account.

**3.** Photoionization of the $4d$ subshell of $Ce@C_{82}^+$ was studied experimentally in [3] and theoretically. Only emission of several (up to three) electrons from the FS was searched in the experiment. The authors of [3] stated that absorption by the encapsulated Ce atom is much smaller than of the isolated one. Here the theoretical investigation becomes more complicated since, contrary to the case of $C_{60}$, $C_{82}$ is non-spherical. To simplify the calculations, we replaced the non-spherical FS by a spherical one with the same volume. For fullerenes shell we choose the same potential parameters as for $C_{60}$. We applied the same RPAE approach as for $Xe@C_{60}$, but the endohedral potential W is of square well type with finite thickness $\Delta$ and inner and outer radius $r_1$ and $r_2 = r_1 + \Delta$, contrary to the $\delta$-type for $C_{60}$ (see [9]). Inside $C_{82}$ cerium gives three electrons to the carbon shell. We assumed, however, that inside $C_{82}$ is $Ce^{+2}$. Having closed shell, $Ce^{+2}$ is much more convenient for theoretical studies than $Ce^{+3}$. The difference between $Ce^{+2}$ and $Ce^{+3}$ is crudely compensated by shifting the theoretical maximum to its experimental value.

The corresponding cross-sections are depicted in Fig.2. To normalize calculated results of calculations to the experimental ones, we used the reduction factor $\approx 0.15$, which is derived from data presented in [3]. Note that in the considered photon energy region $100 eV \leq \omega \leq 180 eV$ the photoionization cross-sections of $Ce^{+2}$ and $Ce^{2+}@C_{82}$ almost coincide [6]. The small reduction factor means that more than $\approx 85\%$ of photoionization acts are accompanied by emission of extra electrons or carbon atoms from the fullerenes shell. As well as in the case of $C_{60}$, even after the fitting of the data for the photoelectron energies of about 20-30 eV, the calculations for photoionization of the caged atom overestimate experimental data at the upper end of the interval., i.e. at the photoelectron energies of about 60-80 eV. Thus some other inelastic channels should be included to obtain the agreement.

---
[1] Atomic system of units $e = m = \hbar = 1$ is used throughout this Letter.

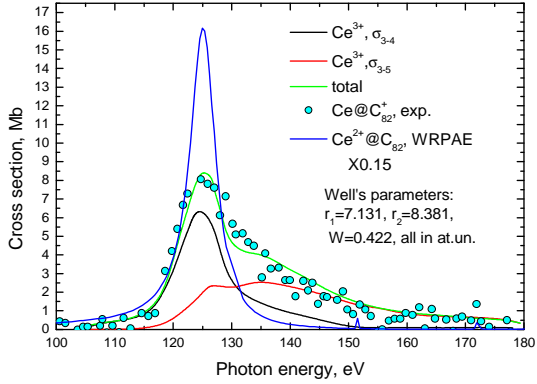

Fig.2. Photoionization cross-sections of $Ce@C_{82}^+$, compared to the cross-section for $Ce^{+++}$ ion. Experiment for $Ce@C_{82}^+$ – from [3]. Our calculation cross-sections for $Ce^{2+}@C_{82}$ are multiplied by the reduction factor 0.15.

**4.** In the measurements of the photoionization cross sections [1-3] the possible energy loss due to excitations of the fullerene shell or emission of several *C* atoms from the fullerene was not taken into account. However, as we have shown recently [7, 8], excitation of the fullerene follows the photoionization of the caged atom with the probability close to unity if the energy of the photoelectron is large enough (but does not reach the values of several *KeV* for the targets considered in this Letter). Below we shall clarify what the words "large enough" mean.

From the first glance the conclusion of [7, 8] looks a little bit surprising, since due to the large radii *R* of the fullerenes shell (FS) $R \gg r_a$ ($r_a$ is the size of the caged atom), the shake-off effects in the FS are small. The same refers to the interaction of the photoelectron with *each* electron of the FS that is determined by the Somerfield parameter

$$\xi^2 = \frac{1}{\upsilon} \approx \frac{13.6 eV}{E} \qquad (2)$$

with *v* being the velocity of the photoelectron in atomic units, and *E* the energy of the photoelectron in *eV*.

The large size *R* and the small width $\Delta$ ($\Delta \ll R$) of the FS enabled to sum the probabilities of the FS excitations, which follows the photoionization of the caged atom:

$$r(E) = 1 - h \exp[-N \ln(1 + \xi^2)]; \quad h = \left| \langle \Phi_0 | \Psi_0 \rangle \right|^2. \qquad (3)$$

Here $h < 1$ is the square of the overlap of the FS wave functions in the ground state with the neutral and ionized caged atom; *N* is the number of the FS electrons, which can participate in an inelastic process in the FS. Applying Eq. (3) to the cases considered in [1-3], we find $r(E) \approx 1$.

Thus the photoelectron loses some part of its energy due to interactions with the FS, and thus no photoelectrons with the energy $E = \omega - I$ will be detected.

However, Eq. (3) was obtained in [7, 8] by employing the closure of the final state FS wave functions. Hence, the derivation requires the energy of the photoelectron to be so large that the most important of FS excitations could be included. In other words, $E$ should be much larger than the energy loss $\varepsilon$ in the fullerene. The photon energies $\omega \approx 100-150 eV$ correspond to $E \approx 80 e$V for the case of $Xe$ considered in [1, 2] and to $E \approx 50 e$V for the case of Ce studied in [3]. For the energies $\varepsilon$ of the excitation of the fullerene, which exceed strongly the FS binding energies $I_{FS}$ (e.g. $I_{FS} \approx 7$ eV for $C_{60}$) the energy distributions drop as $1/\varepsilon^2$. Thus, the values of $\bar{\varepsilon}$ for the valence FS electrons are determined by $I_{FS} \ll \omega \ll E$. The energy distribution at these energies is proportional to $1/\varepsilon^2$ and the energy loss can be estimated as [10, 7]

$$\bar{\varepsilon} \approx \frac{\xi^2 N_v}{4R^2} \ln \frac{E}{I_{FS}}. \tag{4}$$

This leads to the energy loss of about 45 eV for both $C_{60}$ and $C_{82}$. In the case of $Xe@C_{60}^+$ $E$<90 eV, while for $Ce@C_{82}^+$ $E$<80 eV. Thus, employing of closure approximation is not founded rigorously. However, we have seen that the inelastic processes in the FS follow with large probability the photoionization of the caged atom. Thus, the closure approximation employed in [7, 8] includes most important excited states even for relatively small values of the photoelectron energies. In the particular case of $Xe@C_{60}^+$ this is the fragmentation of the FS which loses several atoms of carbon. However, in our analysis we did not need to clarify the nature of the processes with the FS. These can be ejection of several electrons and (or) of several carbon atoms, excitations of collective states, etc.

**5.** A very important consequence of the presented above results is that description of interaction of the photoelectron with the FS by a simple effective potential is not justified even at relatively small photoelectron energies of several dozens of eV. The large role of the inelastic processes prompts that it should be rather an optical potential, similar to that employed in nuclear physics.